\documentclass{article}
\usepackage[latin9]{inputenc}
\usepackage{amsmath}
\usepackage{amssymb}

\makeatletter
\usepackage{amsfonts}

\makeatother

\begin{document}

\title{Proposed physical explanation for the electron spin and related
antisymmetry\\
}
\author{A. M. Cetto, L. de la Peña and A. Valdés\nobreakdash-Hernández \\
%EndAName
Instituto de Física, Universidad Nacional Autónoma de México}
\maketitle

\begin{abstract}
We offer a possible physical explanation for the origin of the electron spin
and the related antisymmetry of the wave function for a two-electron system,
in the framework of nonrelativistic quantum mechanics as provided by linear
stochastic electrodynamics. A consideration of the separate coupling of the
electron to circularly polarized modes of the random electromagnetic vacuum
field, allows to disclose the spin angular momentum and the associated
magnetic moment with a $g$-factor 2, and to establish the connection with
the usual operator formalism. The spin operator turns out to be the
generator of internal rotations, in the corresponding coordinate
representation. In a bipartite system, the distinction between exchange of
particle coordinates (which include the internal rotation angle) and
exchange of states becomes crucial. Following the analysis of the respective
symmetry properties, the electrons are shown to couple in antiphase to the
same vacuum field modes. This finding, encoded in the antisymmetry of the
wave function, provides a physical rationale for the Pauli principle. The
extension of our results to a multipartite system is briefly discussed.

\end{abstract}

\section{Introduction}

Can one speak of the origin of the electron spin? Does it make sense to look
for a physical agent behind the exclusion principle? These two most
prominent features of the electron are customarily taken in nonrelativistic
quantum mechanics as empirical facts. Nevertheless, the concept of spin as a
derived quality rather than an innate one has given rise over the years to a
diversity of suggestive proposals, as is illustrated by our short but
multifarious selection of relevant work \cite{Sch30}-\cite{Jab10}. On the
other hand, despite the various proofs existing in the literature, the
physical gears behind the spin-statistics connection are still unclear (\cite%
{Kap13}-\cite{Kap16} and references therein). Here we address this state of
affairs in an integrated approach using the tools of stochastic
electrodynamics (\textsc{sed}), the theory developed to explain the quantum
behavior of matter as a result of its interaction with the fluctuating
radiation vacuum or zero-point field (\textsc{zpf}) \cite{dlPCetVal15},
taken as a real field. In other words, the purpose of this paper is to
reveal the \emph{physics} behind the spin and its symmetry properties, in
contrast to relativistic quantum mechanics, where the electron spin appears
as a natural element of the formalism.

It should be noted that the electron spin has received some, though rather
limited attention in \textsc{sed}. In particular, by using a
harmonic-oscillator model for the particle and separating the \textsc{zpf}
into components of circular polarization, de la Peña and Jáuregui \cite%
{dlPJau82} obtained the spin angular momentum and the associated magnetic
moment as acquired properties, the former within a numerical factor of order
1. Sachidanandam \cite{Sach83} arrived at similar results for an electron in
a uniform magnetic field. More recently, Muralidhar \cite{Mur12} derived the
electron spin by taking the zero-point energy of the (free) electron as an
energy of rotation within the region of space surrounding the particle.
These various \textsc{sed}-inspired calculations are quite suggestive, in
that they all lead to a result of order $\hbar^{2}$ for the mean square
value of the spin and of order $\hbar$ for the spin projections, exhibiting
the \textsc{zpf} as the source of a kind of (nonrelativistic)
zitterbewegung. This confirms the recurrent proposal of a close relationship
between spin and zitterbewegung, which started with Schrödinger and extends
to our days, as is exemplified in the cited literature\cite{Sch30}-\cite%
{dlPCetVal15}.

Here we use the tools provided by linear stochastic electrodynamics (\textsc{%
lsed}) to tackle the issue of the spin-statistics connection for a
two-electron system, which leads us to propose a specific physical mechanism
for the Pauli exclusion principle, not examined in previous work.{\LARGE {} }%
We start by recalling that according to \textsc{lsed}, the spin appears as
an intrinsic angular momentum of size $\hbar/2$, with an associated magnetic
moment with $g$-factor $g_{S}=2$ \cite{CetdlPVal14}. This paves the way for
the introduction of the corresponding spin-operator formalism and for
establishing contact with its usual (nonrelativistic) quantum treatment. We
then recall that{\LARGE {}\ }two identical particles forming part of the
same system couple to the same mode of the vacuum field; this is the
mechanism behind the entanglement of their state vector, as indicated
previously (\cite{dlPCetVal15}-\cite{ValdlPCet11}). The present work goes
further with respect to Refs. \cite{CetdlPVal14}, \cite{ValdlPCet11}, \cite%
{CetdlP16}, in that here we take advantage of the fact that the \textsc{lsed}
expressions still contain explicitly the relevant field variables describing
the spin orientation, and show that a proper consideration of the rotation
angle associated with the spin angular momentum, combined with a careful
analysis of the exchange properties of the (entangled) state vector, leads
to the antisymmetry of the latter. This result reveals an antiphase coupling
of the two electrons to the common field modes. Since no more than two
particles can couple in antiphase to the same mode, we disclose here the
origin of the Pauli principle. The analysis can be extended to a
multi-electron system, as is briefly discussed in the final part of the
paper.

For a systematic and comprehensive account of \textsc{sed} in its present
form and in particular for a detailed explanation of the emergence of
entangled states, we refer the reader to Refs. \cite{dlPCetVal15} and \cite%
{ValdlPCet11}, respectively. Ref. \cite{CetdlPVal14} contains previous work
on the emergence of spin in \textsc{lsed}. Additionally, some results
presented in this paper draw from previous work on the subject, specifically
Ref. \cite{CetdlP16}.

\section{Intrinsic angular momentum of the vacuum field\label{angle}}

Let us start by recalling from the theory of electric dipole transitions in
atoms that in a single transition the absorbed or emitted radiation carries
one unit of angular momentum, involving an interaction of the electron with
(photonic or external) radiation field modes of circular polarization \cite%
{Sobelman79}. Since in \textsc{sed} we work normally in the electric dipole
approximation, we shall assume that the coupling of electrons to the \textsc{%
zpf} also involves circularly polarized field modes. Accordingly, we expand
the \textsc{zpf} vector potential in terms of plane modes of frequency $%
\omega_{k}=c\left\vert \boldsymbol{k}\right\vert $, wave vector $\boldsymbol{%
k}$, and circular polarization $\gamma$ (see e. g. Ref. \cite{ManWol95},
section 10.6), 
\begin{equation}
\boldsymbol{A}\left(\boldsymbol{r},t\right)=\frac{1}{2}\sqrt{\frac{\hslash}{V%
}}\sum_{\boldsymbol{k},\gamma}\frac{1}{\sqrt{\omega}}\left[-i\boldsymbol{%
\hat{\epsilon}}_{\boldsymbol{k}}^{\gamma}(\varphi)a(\boldsymbol{k}%
,\gamma)e^{i(\boldsymbol{k}\cdot\boldsymbol{r}-\omega t)}+\text{c.c.}\right],
\label{5.1}
\end{equation}
with $V$ the volume of integration. The permittivity $\epsilon_{0}$ of free
space is taken equal to 1, and $\omega=\omega_{k}$. The intensity of the
field is proportional to Planck's constant, which fixes the energy per
normal mode at $\hslash\omega/2$, as corresponds to the zero-point term (see
below).

Let{\large {}{}{}{} }$\boldsymbol{\hat{e}}_{i}${\large {}{}{}{}, }$i=1,2,3,$
be a right-handed triadic of Cartesian unit vectors, with $\boldsymbol{\hat{e%
}}_{3}$ pointing in the direction of $\boldsymbol{k}$; the right and left
circular polarization vectors are then given by $\hat{\epsilon}_{\boldsymbol{%
k}}^{+}(\varphi)=\frac{1}{\sqrt{2}}(\boldsymbol{\hat{e}}_{1}+i\boldsymbol{%
\hat{e}}_{2})e^{i\varphi}${\large {}{}{}{}, }$\hat{\epsilon}_{\boldsymbol{k}%
}^{-}(\varphi)=\frac{i}{\sqrt{2}}(\boldsymbol{\hat{e}}_{1}-i\boldsymbol{\hat{%
e}}_{2})e^{-i\varphi}$, where $\varphi$ is the angle of rotation around the $%
\boldsymbol{\hat{e}}_{3}$-axis. Further, the mode amplitudes $a(\boldsymbol{k%
},\gamma)=e^{i\zeta(\boldsymbol{k},\gamma)}$ vary at random from realization
to realization of the field; this is where stochasticity comes in. For a
maximally incoherent field such as the free \textsc{zpf}, the phases
pertaining to different modes $(\boldsymbol{k},\gamma)$ are statistically
independent. It is convenient to absorb the $\varphi$-factors appearing in
the polarization vectors $\boldsymbol{\hat{\epsilon}}_{\boldsymbol{k}%
}^{\gamma}(\varphi)$ into the mode amplitudes, so that the latter become 
\begin{equation}
a(\boldsymbol{k},\gamma,\varphi)=e^{i\zeta(\boldsymbol{k},\gamma)}e^{i\gamma%
\varphi(\boldsymbol{k},\gamma)},\quad\gamma=\pm1,  \label{a1-1}
\end{equation}
and the polarization vectors reduce to $\boldsymbol{\hat{\epsilon}}_{%
\boldsymbol{k}}^{+}=\frac{1}{\sqrt{2}}(\boldsymbol{\hat{e}}_{1}+i\boldsymbol{%
\hat{e}}_{2}),\boldsymbol{\hat{\epsilon}}_{\boldsymbol{k}}^{-}=\frac{i}{%
\sqrt{2}}(\boldsymbol{\hat{e}}_{1}-i\boldsymbol{\hat{e}}_{2})$. Equation (%
\ref{5.1}) takes then the form 
\begin{equation}
\boldsymbol{A}\left(\boldsymbol{r},t\right)=\frac{1}{2}\sqrt{\frac{\hslash}{V%
}}\sum_{\boldsymbol{k},\gamma}\frac{1}{\sqrt{\omega}}\left[-i\boldsymbol{%
\hat{\epsilon}}_{\boldsymbol{k}}^{\gamma}a(\boldsymbol{k},\gamma,%
\varphi)e^{i(\boldsymbol{k}\cdot\boldsymbol{r}-\omega t)}+\text{c.c.}\right].
\label{5.1a}
\end{equation}

Integrating over the entire volume, with the help of Eq. (\ref{a1-1}), one
readily obtains for every mode contained in (\ref{5.1a}) a fixed (nonrandom)
energy $H_{\boldsymbol{k}}^{\gamma}=\hslash\omega/2=\hslash\omega_{%
\boldsymbol{k}}/2,$ a nonrandom linear momentum $\boldsymbol{P}_{\boldsymbol{%
k}}^{\gamma}=\hslash\omega\boldsymbol{\hat{\mathbf{k}}}/2c,$ and a fixed
intrinsic angular momentum along the direction of propagation $\boldsymbol{J}%
_{k}^{\gamma}$ of value 
\begin{equation}
\boldsymbol{~J}_{\boldsymbol{k}}^{\gamma}=\int_{V}\left(\boldsymbol{E}\times%
\boldsymbol{A}\right)_{\boldsymbol{k}}^{\gamma}d^{3}r=\gamma\frac{\hslash}{2}%
\boldsymbol{\hat{\mathbf{k}},}\quad\gamma=\pm1.  \label{10}
\end{equation}
These values coincide with the results reported in the literature \cite%
{ManWol95}. Since there are as many modes in the $\boldsymbol{k}$ direction
as in the $-\boldsymbol{k}$ direction, the total linear momentum vanishes
for every $\omega.$ Further, for the free field the contributions of the two
polarizations compensate each other for every $\boldsymbol{k}$, so that the
total intrinsic angular momentum vanishes as well. This may explain why, in 
\textsc{sed} as well as in \textsc{qed}, these terms are normally omitted.
Yet every \emph{individual} field mode $(\boldsymbol{k},\gamma)$ does have
an \emph{intrinsic} angular momentum of value $\pm\hslash/2$ along the
direction of propagation, according to Eq. (\ref{10}). This decomposition of
the radiation field into orbital and intrinsic (spinorial) components is
legitimated by the fact that for a free electromagnetic field the intrinsic
angular momentum is a constant of the motion; a detailed discussion of this
matter can be seen in Ref. \cite{ManWol95}, Section 10.6.3.

\section{Intrinsic angular momentum of the particle\label{spin}}

For an analysis of the effect of the vacuum field on the angular momentum of
the electron, the approach provided by \textsc{lsed} is particularly
convenient. As shown elsewhere (Ref. \cite{dlPCetVal15} and references
therein) the theory furnishes a description of the stationary states of the
mechanical system once it has reached the quantum regime \textemdash i.e.,
when the system has acquired ergodic properties and detailed energy balance
has been attained between particle and vacuum field. Additional effects of
the radiation terms are then negligible and may be omitted in the
radiationless approximation. The ensuing description is, formally, entirely
equivalent to the Heisenberg quantum description; yet it still contains
relevant information about those field modes that play a central role in
sustaining the stationary states. This information also plays a key role in
the analysis that follows.

Let us consider a charged, pointlike particle subject to an external
conservative force, typically an atomic electron. According to \textsc{lsed}%
, a generic dynamical variable $G(t)$ pertaining to the particle in a
stationary state $\alpha$ has the form 
\begin{equation}
G_{\alpha}(t)=\tilde{G}_{\alpha\alpha}+\sum_{\beta\neq\alpha}\tilde{G}%
_{\alpha\beta}a_{\alpha\beta}e^{i\omega_{\alpha\beta}t},  \label{14'}
\end{equation}
where the index $\beta\neq\alpha$ represents any other stationary state; the
set $\left\{ \alpha\beta\right\} $ depends on the specific problem. The $%
\tilde{G}_{\alpha\beta}$ turn out to be the matrix elements (in the energy
representation) of the respective quantum operator $\hat{G}$, so that $%
\tilde{G}_{\alpha\alpha}$ is the expectation value of $G$ in state $\alpha$.
The field amplitudes $a_{\alpha\beta}$ pertain to those modes of the field
(of frequency $\omega_{\alpha\beta}$) to which the particle responds
resonantly, and are given by (cf. Eq. (\ref{a1-1})) 
\begin{equation}
a_{\alpha\beta}(\varphi)=e^{i\zeta_{\alpha\beta}}e^{i\gamma_{\alpha\beta}%
\varphi},\quad\gamma_{\alpha\beta}=\pm1.  \label{a2}
\end{equation}
As manifested by the dependence of $G_{\alpha}$ on the $a_{\alpha\beta}$,
the particle variables are driven linearly, so to say, by such field modes
(under stationarity neither $\tilde{G}_{\alpha\beta}$ nor $%
\omega_{\alpha\beta}$ depend on the stochastic amplitudes $a_{\alpha\beta}$%
). A notable feature of the quantum regime is that when ergodicity is
imposed, the phases of the $a_{\alpha\beta}$ become correlated in such a way
that the \emph{chain rule} $a_{\alpha\beta^{\prime}}a_{\beta^{\prime}%
\beta}=a_{\alpha\beta}$ holds (no summation over repeated indices) \cite%
{dlPCetVal15}, with $\beta^{\prime}$ any stationary state, which implies $%
\zeta_{\alpha\beta}=\zeta_{\alpha}-\zeta_{\beta}$, $\omega_{\alpha\beta}=%
\omega_{\alpha}-\omega_{\beta}$, and $\gamma_{\alpha\beta}=\gamma_{\alpha}-%
\gamma_{\beta}$. Physically this is a result of the effect of the radiating
particle on the field under stationary conditions; mathematically, this
guarantees that the product of two (or more) dynamical variables can also be
written as a linear expansion of the form (\ref{14'}). Further, the relation 
$\omega_{\alpha}=\mathcal{E}_{\alpha}/\hbar$ is shown to hold, with $%
\mathcal{E}_{\alpha}$ the energy associated with state $\alpha$, meaning
that the resonance frequencies are just the transition frequencies as given
by Bohr's rule \cite{dlPCetVal15}.

Equation (\ref{14'}) applies in particular to the components of $\boldsymbol{%
x}(t)$ and $\boldsymbol{p}(t)=m\dot{\boldsymbol{x}}(t)$ in state $\alpha$.
The average value of the angular momentum component $L_{z}=xp_{y}-yp_{x}$
becomes thus 
\begin{equation}
\left\langle L_{z}\right\rangle _{\alpha}=\left\langle \alpha\right\vert 
\hat{L}_{z}\left\vert \alpha\right\rangle =im\sum\nolimits
_{\beta}\omega_{\beta\alpha}\left(\tilde{x}_{\alpha\beta}\tilde{y}%
_{\beta\alpha}-\tilde{y}_{\alpha\beta}\tilde{x}_{\beta\alpha}\right),
\label{16}
\end{equation}
with $\tilde{x}_{\alpha\beta}=\left\langle \alpha\right\vert \hat{x}%
\left\vert \beta\right\rangle ,$ $\tilde{y}_{\alpha\beta}=\left\langle
\alpha\right\vert \hat{y}\left\vert \beta\right\rangle $, and $\left\vert
\alpha\right\rangle $ a vector in $\mathcal{H},$ the corresponding Hilbert
space of states.

Since the electron is driven by the (electric component of the) circularly
polarized \textsc{zpf} modes, we shall consider separately the two circular
polarizations of the field modes contributing to $\left\langle
L_{z}\right\rangle _{\alpha}$, which are those that propagate along the $z$
axis. It is therefore convenient to use cylindrical variables for the
particle: $x^{+}=\boldsymbol{x}\mathbf{\cdot}\boldsymbol{\hat{\epsilon}}^{+}=%
\frac{1}{\sqrt{2}}(x+iy),\enskip x^{-}=\boldsymbol{x}\mathbf{\cdot}%
\boldsymbol{\hat{\epsilon}}^{-}=\frac{1}{\sqrt{2}}(ix+y)$, and $x^{k}=z$. In
terms of these, (\ref{16}) rewrites as 
\begin{equation}
\left\langle L_{z}\right\rangle _{\alpha}=m\sum\nolimits
_{k}\omega_{\beta\alpha}\left(\left\vert \tilde{x}_{\alpha\beta}^{+}\right%
\vert ^{2}-\left\vert \tilde{x}_{\alpha\beta}^{-}\right\vert ^{2}\right).
\label{19}
\end{equation}
To calculate the separate terms in (\ref{19}) we resort to the commutator $%
\left[\hat{x},\hat{p}\right]=i\hbar,$ which in \textsc{lsed} is shown to be
a further consequence of the particle-field interaction in the quantum
regime.\cite{CetdlP16} In cylindrical variables it takes the form 
\begin{equation}
\hbar=m\sum\nolimits _{\beta}\omega_{\beta\alpha}\left(\left\vert \tilde{x}%
_{\alpha\beta}^{+}\right\vert ^{2}+\left\vert \tilde{x}_{\alpha\beta}^{-}%
\right\vert ^{2}\right).  \label{20}
\end{equation}
By combining with Eq. (\ref{19}) we get 
\begin{equation}
\left\langle L_{z}\right\rangle _{\alpha}=\left\langle M_{z}\right\rangle
_{\alpha}^{+}+\left\langle M_{z}\right\rangle _{\alpha}^{-},  \label{22}
\end{equation}
with 
\begin{equation}
\left\langle M_{z}\right\rangle _{\alpha}^{+}=\tfrac{1}{2}\left\langle
L_{z}\right\rangle _{\alpha}+\tfrac{1}{2}\hbar,\quad\left\langle
M_{z}\right\rangle _{\alpha}^{-}=\tfrac{1}{2}\left\langle L_{z}\right\rangle
_{\alpha}-\tfrac{1}{2}\hbar.  \label{23}
\end{equation}
Notice that even if (and when) $\left\langle L_{z}\right\rangle _{\alpha}$
is zero, the separate contributions to $\left\langle M_{z}\right\rangle
_{\alpha}^{+}$ and $\left\langle M_{z}\right\rangle _{\alpha}^{-}$ do not
vanish. Each one contains one-half of the orbital angular momentum along the 
$z$ direction plus an intrinsic angular momentum component $\pm\hbar/2$
associated with either one or the other polarization; this intrinsic term we
identify with the spin component $S_{z}$. By summing over the polarizations,
the spin terms cancel each other and we are left with the orbital term only.
So even though its contributions are concealed by the summation, the spin
emerges as manifestation of the coupling of the particle to the separate
polarized field modes of the \textsc{zpf}.

To express $\left\langle M_{z}\right\rangle _{\alpha}^{\pm}$ as the average
of an operator $\hat{M}_{z}$, we note that $\left\langle L_{z}\right\rangle
_{\alpha}$ does not depend on the spin state and the latter does not depend
on the atomic state. We therefore decompose $\mathcal{H}$ as $\mathcal{H}=%
\mathcal{H}_{0}\otimes\mathcal{H}_{S}$, with $\mathcal{H}_{0}$ the non-spin
Hilbert space, spanned by the orbital state vectors, denoted henceforth by $%
\left\vert \alpha_{0}\right\rangle $, and $\mathcal{H}_{S}$ a bidimensional
space spanned by the orthonormal vectors $\left\{ \left\vert
\sigma\right\rangle \right\} =(\left\vert +\right\rangle ,\left\vert
-\right\rangle )$ representing the eigenstates of a spin operator which we
call $\boldsymbol{\hat{S}}$. In terms of $\left\vert \alpha\right\rangle
=\left\vert \alpha_{0},\sigma\right\rangle =\left\vert
\alpha_{0}\right\rangle \otimes\left\vert \sigma\right\rangle $, Eqs. (\ref%
{23}) become 
\begin{equation}
\left\langle \alpha\right\vert \hat{M}_{z}\left\vert \alpha\right\rangle =%
\tfrac{1}{2}\left\langle \alpha_{0}\right\vert \hat{L}_{z}\left\vert
\alpha_{0}\right\rangle +\left\langle \sigma\right\vert \hat{S}%
_{z}\left\vert \sigma\right\rangle ,  \label{24}
\end{equation}
with $\sigma=\pm1.$ In terms of the Pauli matrix $\hat{\sigma}_{z}$ we have $%
\hat{S}_{z}=$$\hbar\hat{\sigma}_{z}/2$, and (\ref{24}) becomes 
\begin{equation}
\left\langle \alpha\right\vert \boldsymbol{\hat{M}}\cdot\boldsymbol{\hat{z}}%
\left\vert \alpha\right\rangle =\left\langle \alpha_{0}\sigma\right\vert
\left(\tfrac{1}{2}\boldsymbol{\hat{L}}+\boldsymbol{\hat{S}}\right)\cdot%
\boldsymbol{\hat{z}}\left\vert \alpha_{0}\sigma\right\rangle ,  \label{26}
\end{equation}
where 
\begin{equation}
\boldsymbol{\hat{S}}=\tfrac{1}{2}\hbar\boldsymbol{\hat{\sigma}}.  \label{28}
\end{equation}
The independence of $\langle\hat{L}_{z}\rangle$ from $\sigma$ and of $\langle%
\hat{S}_{z}\rangle$ from $\alpha_{0}$, indicates that under the present
conditions, the fluctuations associated with the spin (a non-relativistic
zitterbewegung, taken as \emph{internal}) are not correlated with those that
characterize the mean instantaneous kinematics of the particle. This is a
characteristic nonrelativistic independence. (The spaces of the two angular
momenta may of course become connected by the presence of magnetic forces.)
By \emph{internal }we are referring to the jiggling \emph{around} the local
mean position of the particle; when a translational motion is superimposed,
this corresponds to a kind of helicoidal motion.

Along with energy and momentum, a given mode of the \textsc{zpf} is thus
seen to also transfer a \textit{minimum} angular momentum to the particle,
equal to its mean spin value $\hbar/2$, independent of the binding force. It
should be stressed that this ineluctable angular momentum does not refer to
a spinning motion of the (pointlike!) particle, but rather to a rotation
around its instantaneous position along its (comparatively smooth)
trajectory.\footnote{%
In Ref. \cite{CetdlPVal14}, this rotational motion is shown to be generated
by the torque due to the electric component of the \textsc{zpf}; see Eq.
(41) and the related discussion in that paper.} This additional motion
endorses the notion frequently encountered (already in Schrödinger, Ref. 
\cite{Sch30}) that the spin has its origin just in the zitterbewegung. At
the same time it explains why the spin cannot be associated with the
instantaneous mean local coordinates of the particle, since it represents a
motion \textit{around} the latter. Further, it elucidates the reason for the
presence of spin in Dirac's equation, since this (relativistic) theory
predicts the zitterbewegung, a phenomenon totally ignored by contrast in the
Schrödinger theory, where instead it needs to be introduced by hand.

With these elements we can readily calculate the $g_{S}$-factor associated
with the electron spin \textemdash which in the nonrelativistic case is also
normally introduced by hand. For this purpose consider an atomic electron
acted on by a static uniform magnetic field $\boldsymbol{B}=B\boldsymbol{%
\hat{z}}$ in addition to the binding Coulomb force. The additional
contribution to the Hamiltonian is $\hat{H}=-\boldsymbol{\hat{\mu}}\cdot%
\boldsymbol{B},$ with $\boldsymbol{\hat{\mu}}=-(g_{L}\mu_{0}\boldsymbol{\hat{%
L}})/\hbar$, $\mu_{0}=\left\vert e\right\vert \hbar/(2mc)$ the Bohr magneton
($e=-\left\vert e\right\vert $), and $g_{L}=1$. The corresponding mean
energy $\mathcal{E}=\mu_{0}B\langle\hat{L}_{z}\rangle/\hbar$ can be
separated using Eqs. (\ref{22}) and (\ref{23}) into $\mathcal{E}%
^{\pm}=\mu_{0}B\left(\langle\hat{L}_{z}\rangle+2\langle\hat{S}%
_{z}\rangle^{\pm}\right)/2\hbar.$ The partial Hamiltonians describing the
magnetic interaction of the electron with right and left modes are therefore 
$\hat{H}_{LS}^{+}=\hat{H}_{LS}^{-}=\mu_{0}B\left(\hat{L}_{z}+2\hat{S}%
_{z}\right)/2\hbar$, and the full Hamiltonian reads 
\begin{equation}
\hat{H}_{LS}=\hat{H}_{LS}^{+}+\hat{H}_{LS}^{-}=\frac{\mu_{0}}{\hbar}B\left(%
\hat{L}_{z}+2\hat{S}_{z}\right),  \label{35}
\end{equation}
with the correct $g$$_{S}$-factor of $2$ for the spin magnetic moment. It is
clear from this derivation that the value of $g$$_{S}$ is linked with the
two polarizations of the \textsc{zpf}.\cite{dlPJau82} By writing Eq. (\ref%
{35}) as $\hat{H}_{LS}=-\boldsymbol{\hat{\mu}}\cdot\boldsymbol{B}$, the
operator $\boldsymbol{\boldsymbol{\hat{M}}}$ turns out to be proportional to
the total magnetic moment operator $\boldsymbol{\hat{\mu}}$ of the atomic
electron, 
\begin{equation}
\boldsymbol{\hat{\mu}}=-\frac{\mu_{0}}{\hbar}(\boldsymbol{\hat{L}}+2%
\boldsymbol{\hat{S}})=-\frac{2\mu_{0}}{\hbar}\boldsymbol{\boldsymbol{\hat{M}}%
}.  \label{37}
\end{equation}

Now we reformulate the state vectors $\left\vert \alpha\right\rangle
=\left\vert \alpha_{0},\sigma\right\rangle =\left\vert
\alpha_{0}\right\rangle \otimes\left\vert \sigma\right\rangle $ considering
the internal rotation angle $\varphi$ introduced in section \ref{angle}. As
Eq. (\ref{14'}) indicates, $G_{\alpha}$ depends on $\varphi$ through the
amplitudes (\ref{a2}), with $\gamma_{\alpha\beta}=\gamma_{\alpha}-\gamma_{%
\beta}=\pm1,$ as explained after Eq. (\ref{a2}). We profit from the
structure of Eq. (\ref{14'}) to transfer the dependence on $\varphi$ to the
state vectors $\left\vert \alpha\right\rangle $, and define 
\begin{equation}
\left\vert \alpha(\varphi)\right\rangle
=e^{i\gamma_{\alpha}\varphi}\left\vert \alpha_{0},\sigma\right\rangle ,
\label{144}
\end{equation}
meaning that the angle $\varphi$ is now associated with the particle.

To determine the set of values $\left\{ \gamma_{\alpha}\right\} $, let us
assume that there exist (at least) three different possible values, say $%
\gamma_{\alpha},\gamma_{\beta}$, and $\gamma_{\delta}$. Then $%
\gamma_{\alpha}-\gamma_{\delta}=\pm1$ and $\gamma_{\beta}-\gamma_{\delta}=%
\pm1$ must hold simultaneously. From the latter it follows that $%
\gamma_{\delta}=1+\gamma_{\beta}\mp1$, which gives for $\gamma_{\delta}$ the
values $\gamma_{\beta}$ or $\gamma_{\beta}+2,$ contrary to $%
\gamma_{\delta}=\gamma_{\beta}\mp1;$ hence $\gamma_{\alpha}$ is a
dichotomous parameter, like $\sigma$. One of its values can be made to refer
to polarization $+1$, the other to $-1,$ the case of equal values being
excluded by $\gamma_{\alpha}-\gamma_{\beta}=\pm1$. They differ then in sign,
so that $\gamma_{\beta}=-\gamma_{\alpha}=\gamma_{\alpha}\pm1,$ or $%
\gamma_{\alpha}=\pm1/2.$ This leads us to identify the values of the
parameter $\gamma_{\alpha}$ with the eigenvalues $\pm1/2$ of the spin
projection in units of $\hbar$. Notice further that the vectors (\ref{144})
become then eigenfunctions of the operator $-i\partial_{\varphi}$ with
eigenvalues $\pm1/2$, whence in this $\varphi$-representation \textemdash $%
\varphi$ being the \emph{internal} rotation angle\textemdash{} the spin
operator $\hat{S}_{z}$ becomes 
\begin{equation}
\hat{S}_{z}=-i\hbar\partial_{\varphi},  \label{144'}
\end{equation}
in analogy with the orbital angular momentum operator \emph{$\hat{L}%
_{z}=-i\hbar\partial_{\phi}.$}

\section{Symmetry properties of the bipartite state vector}

We recall that in nonrelativistic quantum mechanics the antisymmetry of the
state vector for fermions is normally postulated, or borrowed from
relativistic quantum field theory. As mentioned above, although there exist
several proposed quantum-mechanical derivations of the spin-symmetry
connection, the physical reasons for this connection remain unascertained
(see Refs. \cite{Kap13} and \cite{Kap16} for a discussion).

In the following we will analyse the spin-symmetry connection by applying
the tools of \textsc{lsed} to a stationary state of a two-electron system.
This will prove to have the advantage of resorting to a more complete
description of the state of the bipartite system, which involves the
relevant field variables and the internal rotation angles in addition to the
variables that represent the (quantum) states of the two particles according
to the usual description. The symmetry properties of the bipartite state
vector will therefore be determined by considering the exchange of the \emph{%
full} set of variables.

A major outcome of \textsc{lsed} is that any stationary state of a system of
two identical particles becomes described in terms of an entangled state
vector of the form (\cite{dlPCetVal15}, \cite{ValdlPCet11}) 
\begin{equation}
\left\vert \psi\right\rangle _{12}^{AB}=\tfrac{1}{\sqrt{2}}\Big(\left\vert
A\right\rangle +\lambda_{AB}\left\vert B\right\rangle \Big),  \label{145}
\end{equation}
with the entanglement parameter $\lambda_{AB}$ given by the product of the
random field amplitudes, the subindices referring to particles 1 and 2, 
\begin{equation}
\lambda_{AB}=\left(e^{i\zeta_{\alpha\alpha^{\prime}}}\right)_{1}\,\left(e^{i%
\zeta_{\alpha^{\prime}\alpha}}\right)_{2}.  \label{B13-1-1}
\end{equation}
In (\ref{145}), state $\left\vert A\right\rangle $ is given by 
\begin{equation}
\left\vert A\right\rangle =\left\vert \alpha(\varphi_{1})\right\rangle
_{1}\left\vert \alpha^{\prime}(\varphi_{2})\right\rangle
_{2}=e^{-i\sigma\varphi_{1}}\left\vert \alpha_{0},\sigma\right\rangle
e^{-i\sigma^{\prime}\varphi_{2}}\left\vert
\alpha_{0}^{\prime},\sigma^{\prime}\right\rangle ,  \label{B19'}
\end{equation}
and similarly for $\left\vert B\right\rangle ,$ with $\alpha,\alpha^{\prime}$
and $\sigma,\sigma^{\prime}$ interchanged; $A(\alpha,\alpha^{\prime})$, $%
B(\alpha^{\prime},\alpha)$, with $\alpha^{\prime}\neq\alpha$, are different
composite states having the same total energy $\mathcal{E}_{A}=\mathcal{E}%
_{\alpha}+\mathcal{E}_{\alpha^{\prime}}=\mathcal{E}_{B}$. In expressions
such as (\ref{B19'}), the first state vector refers always to particle 1 and
the second one to particle 2. As is clear from (\ref{B13-1-1}), the
entanglement results from the fact that identical particles couple through
common relevant \textsc{zpf} modes, which allows for the emergence of
correlations between them.

Let us analyse the symmetry properties of the state vector $\left\vert
\psi\right\rangle _{12}^{AB}$ under different exchange operations. The
expression (\ref{145}) lends itself to two such operations: one can either
exchange states $A$ and $B$, or exchange particles 1 and 2. In contrast to
quantum mechanics, these operations are \emph{not} equivalent because they
act distinctly either on the particles (which are identical) or on the
states (which are different, by construction). In particular, the exchange
of particles involves the internal angular coordinate $\varphi$ (see below),
a variable foreign to the usual quantum description.

The first operation $(A,B)\rightarrow(B,A)$ leads to 
\begin{equation}
\left\vert \psi\right\rangle _{12}^{BA}=\tfrac{1}{\sqrt{2}}\Big(\left\vert
B\right\rangle +\lambda_{BA}\left\vert A\right\rangle \Big).  \label{145b}
\end{equation}
According to (\ref{B13-1-1}), $\lambda_{BA}=\left(e^{i\zeta_{\alpha^{\prime}%
\alpha}}\right)_{1}\,\left(e^{i\zeta_{\alpha\alpha^{\prime}}}\right)_{2}=%
\lambda_{AB}^{*}=\lambda_{AB}^{-1},$ whence from (\ref{145}) and (\ref{145b}%
), 
\begin{equation}
\left\vert \psi\right\rangle _{12}^{BA}=\lambda_{AB}^{-1}\left\vert
\psi\right\rangle _{12}^{AB}.  \label{145bb}
\end{equation}
Since $\left\vert A\right\rangle $ and $\left\vert B\right\rangle $
represent two different states, in principle $\left\vert \psi\right\rangle
_{12}^{BA}$ may be different from $\left\vert \psi\right\rangle _{12}^{AB}$.

Now we perform an exchange of particles $(1,2)\rightarrow(2,1)$. In contrast
to the exchange of states, this one should have no effect on the composite
state vector since the particles are identical, i.e., 
\begin{equation}
\left\vert \psi\right\rangle _{21}^{AB}=\left\vert \psi\right\rangle
_{12}^{AB}.  \label{42}
\end{equation}
For this operation we need to consider explicitly the dependence of $%
\left\vert \psi\right\rangle _{12}^{AB}$ on the intrinsic (internal) angular
coordinate $\varphi$. As discussed in Ref. \cite{Jab10}, one must take care
that the rotations that take particle 1 to the azimuthal angle of particle 2
and vice versa, are both made in the same sense \textemdash say clockwise.
When $\varphi_{2}>\varphi_{1}$, $\varphi_{1}$ transforms into $\varphi_{2}$
and $\varphi_{2}$ transforms into $2\pi+\varphi_{1}$, so that one gets 
\begin{equation}
\left\vert \psi\right\rangle _{21}^{AB}=\tfrac{1}{\sqrt{2}}\Big(e^{-2\pi
i\sigma^{\prime}}\left\vert B\right\rangle +\lambda_{BA}e^{-2\pi
i\sigma}\left\vert A\right\rangle \Big)=-\tfrac{\lambda_{BA}}{\sqrt{2}}\Big(%
\lambda_{AB}\left\vert B\right\rangle +\left\vert A\right\rangle \Big),
\label{43}
\end{equation}
since $e^{-i2\pi\sigma}=e^{-i2\pi\sigma^{\prime}}=-1$ and $%
\lambda_{AB}\lambda_{BA}=1$. When $\varphi_{2}<\varphi_{1}$, $\varphi_{2}$
transforms into $\varphi_{1}$, $\varphi_{1}$ transforms into $%
2\pi+\varphi_{2}$, and the exchange applied to $\left\vert \psi\right\rangle
_{12}$ gives again Eq. (\ref{43}). Therefore we have in both cases 
\begin{equation}
\left\vert \psi\right\rangle _{21}^{AB}=-\lambda_{BA}\left\vert
\psi\right\rangle _{12}^{AB}.  \label{45}
\end{equation}
By comparing this result with Eq. (\ref{42}) we obtain

\begin{equation}
\lambda_{AB}=-1,  \label{46}
\end{equation}
which introduced into (\ref{145}) or (\ref{145bb}) leads to the well-known
antisymmetric form of the state vector 
\begin{equation}
\left\vert \psi\right\rangle _{12}^{AB}=\tfrac{1}{\sqrt{2}}\Big(\left\vert
A\right\rangle -\left\vert B\right\rangle \Big)=-\left\vert
\psi\right\rangle _{12}^{BA},  \label{49}
\end{equation}
indicating that the permutation of fermion states $A\leftrightarrow B$
produces an overall change of sign, just as in quantum theory.

The above outcome has a further important physical implication: from (\ref%
{B13-1-1}) we note that $\lambda_{AB}=-1$ implies $\left(e^{i\zeta_{\alpha%
\alpha^{\prime}}}\right)_{1}=e^{\pi
i}\left(e^{i\zeta_{\alpha\alpha^{\prime}}}\right)_{2}$, indicating that the
coupling of particles 1 and 2 to the mode of frequency $\omega_{\alpha%
\alpha^{\prime}}$ occurs \emph{out of phase}, with a phase difference of $\pi
$. For particles with symmetric wave functions (i. e. with $\lambda_{AB}=1$%
), by contrast, the coupling occurs \emph{in phase}, as seen from (\ref%
{B13-1-1}).\footnote{%
Note that also in this latter case the assumption that the two-particle
state vector does not change sign upon particle exchange, holds true.} While
an arbitrary number of (identical) particles can couple in phase to a single
mode, no more than two particles can couple with a phase difference of $\pi$
to the same mode. This throws a new light on the Pauli exclusion principle.

\section{Final discussion}

The above analysis can be extended to a multielectron system, subject again
to a common \textsc{zpf}, thanks to the fact that the chain rule discussed
in Sect. \ref{spin} remains in force for an arbitrary product of spin
phases. To determine the resulting state of the system one must consider the
various possible configurations $A,B,C,$ $...$ of stationary states
corresponding to the same total energy $\mathcal{E}_{A}=\mathcal{E}_{B}=%
\mathcal{E}_{C}=...$ Yet direct application of this procedure becomes rather
cumbersome, as the degeneracy increases rapidly with the number of
particles. A convenient approach is to consider first any pair of electrons
of the system, say those in states $\alpha,\alpha^{\prime}$. By taking
successively every possible pair, all relevant frequencies will be accounted
for, and all the respective symmetries will thus be included. Since as a
result no pair of electrons can be in the same (single-particle) state, the
state of the entire system will be described by a totally antisymmetric,
multiply entangled state vector built of different bipartite single-particle
states that carry the factor $(-1)^{2p\sigma}=(-1)^{p}$ in front of each
term, where $p$ stands for the number of transpositions in the permutation
needed to reach the corresponding exchanged state, starting from the initial
state.

The calculations presented here confirm the physical picture of the spin of
the electron as a helicoidal motion (a zitterbewegung) around the local mean
trajectory, adding an effective structure to the originally pointlike
particle. The operator $-i\hbar\partial_{\varphi}$ turns out to be the
generator of internal rotations, according to Eq. (\ref{144'}). Further, for
a bi- or multielectron system, consideration of the permanent presence of
the background field resonantly connecting the particles serves to unveil
the persistent mystery of the physics behind the spin-statistics
association. The mediation of the \textsc{zpf} turns out to have a definite
role in fixing the statistics. Nevertheless, the above arguments appear so
far to be insufficient to deal with the universe of bosonic particles, which
in general are hadrons and subject also to nuclear interactions; other
mechanisms most likely intervene in the definition of the total exchange
symmetry properties of such compound systems.

\emph{Acknowledgment}\textit{s.} The authors acknowledge helpful comments
and suggestions from an anonymous referee. This work was supported in part
by DGAPA-UNAM, through project IN104816.

\end{document}